\documentclass[prd,aps,10pt,twocolumn,showpacs,showkeys,reprint,nofootinbib]{revtex4}
 
\usepackage[dvips]{graphicx}
\usepackage{amssymb}
\usepackage{filecontents}
\usepackage{amsmath}
\usepackage{color}
\usepackage{float}
\RequirePackage[colorlinks,citecolor=blue,urlcolor=blue,linkcolor=blue]{hyperref}
\def\pmb#1{\setbox0=\hbox{$#1$}%
  \kern-.025em\copy0\kern-\wd0
  \kern.05em\copy0\kern-\wd0
  \kern-.025em\raise.0433em\box0}

\def\beq{\begin{equation}}
\def\eeq{\end{equation}}
\begin{document}
\title{Relativistic Tidal Accelerations in the Exterior Schwarzschild Spacetime}

\author{Mahmood \surname{Roshan}$^{1,2}$}
\email{mroshan@um.ac.ir}
\author{Bahram \surname{Mashhoon}$^{2,3}$}
\email{mashhoonb@missouri.edu}

\affiliation{$^1$Department of Physics, Faculty of Science, Ferdowsi University of Mashhad, P.O. Box 1436, Mashhad, Iran \\
$^2$School of Astronomy, Institute for Research in Fundamental Sciences (IPM), P. O. Box 19395-5531, Tehran, Iran\\
$^3$Department of Physics and Astronomy, University of Missouri, Columbia, Missouri 65211, USA\\
}

\begin{abstract}
We study further a general relativistic mechanism for the acquisition of tidal energy by free test particles near a gravitationally collapsed configuration. Specifically, we investigate the solutions of timelike geodesic equation in a Fermi normal coordinate system established about the world line of an accelerated observer that remains spatially at rest in  the exterior Schwarzschild spacetime. Such static observers in effect define the rest frame of the collapsed source. The gain in tidal energy is due to local spacetime curvature. Previous work in this direction in connection with astrophysical jets  involved geodesic motion along the Kerr rotation axis where outward moving particles could be tidally accelerated to almost the speed of light if their initial speed, as measured by the reference observer, is above a certain threshold escape velocity. We focus here for simplicity on the exterior Schwarzschild spacetime and show that the effective tidal acceleration to near the speed of light can occur for geodesic motion in certain radially outward directions.  The implications of these results for the physics of gravitationally collapsed systems are briefly discussed.
\end{abstract}

\maketitle

\section{Introduction}

Half a century ago, Wheeler introduced the idea that the gravitational tidal forces of black holes may play a significant role in jet formation in the nuclei of active galaxies~\cite{Wheeler}. Wheeler's pioneering work has been followed by many subsequent theoretical studies. Of these, some have focused on the dynamics of astrophysical jets and the physics of black holes; see, for instance, Refs.~\cite{Blandford:1977ds, Blandford:1982di, Felice, Gariel:2007st, Gariel:2016vql, Zhang:2017kmq, Poirier:2015cyu, Poirier:2016rul,  Tucker:2016wvt, Tucker:2018xle} and the references cited therein. Some other studies, however,  have concentrated on relativistic tidal phenomena in black hole spacetimes; see, for example, Refs.~\cite{mas77, Marck, MaMc, Chicone:2002kb, Kojima:2005dm, Bini:2007zzb, Bini:2012zze, Bini:2017uax, Junior:2020yxg, Junior:2020par, Mashhoon:2020tha} and the references therein. 

In the present work, we study further a recent theoretical tidal acceleration mechanism along the rotation axis of Kerr spacetime~\cite{Bini:2017uax}. It is of basic  interest in connection with energetic outflows that is characteristic of astrophysical jets. Briefly, we consider an astrophysical object of mass $M$ and angular momentum $J$ whose gravitational field is described by the Kerr metric, which we represent in standard Boyer-Lindquist coordinates $(\hat{t},r,\theta,\phi)$ as~\cite{Chandra}
\begin{eqnarray}
\label{I0}
-ds^2&=&-d\hat{t}^2+\frac{\Sigma}{\Delta}\,dr^2+\Sigma\, d\theta^2 +(r^2+\hat{a}^2)\sin^2\theta\, d\phi^2\nonumber\\
&& +\frac{2Mr}{\Sigma}\,(d\hat{t}-\hat{a}\,\sin^2\theta\, d\phi)^2\,.
\end{eqnarray}
Here, $\hat{a} = J/(Mc)$ is the specific angular momentum parameter, $\Sigma=r^2+\hat{a}^2\cos^2\theta$ and $\Delta=r^2-2Mr+\hat{a}^2$. Next, we imagine the set of test observers that remain spatially at rest in the exterior Kerr spacetime; indeed, these accelerated observers in effect define the rest frame of the gravitational source. The source is a Kerr \emph{black hole} for  $\hat{a}\le M$; however, this circumstance has no essential impact on our basic physical considerations here. Let us now consider a test observer that is at rest at a radial coordinate $r$ along the rotation axis of the exterior Kerr spacetime. We establish an approximate Fermi normal coordinate system with coordinates $(T, X, Y, Z)$ along the natural nonrotating orthonormal tetrad frame adapted to this observer~\cite{Synge}.  The test observer permanently occupies the spatial origin of this Fermi system; thus, $(T, 0, 0, 0)$ are the Fermi coordinates of the test observer and $T$ is its proper time. With respect to this test observer and in terms of approximate Fermi coordinates established along its world line, free test particles starting out from the origin of the Fermi system and moving outward along the rotation axis (i.e., the $X$ axis in Fermi coordinates) with initial speed $V_0$, $0 < V_0 < 1$, turn around and fall back toward the gravitational source if $V_0$ is below a certain threshold given approximately by $[GM/(2\,r)]^{1/2}$. However, above this threshold, free test particles can eventually get tidally accelerated to almost the speed of light~\cite{Bini:2017uax}. That is, a free test particle with unit 4-velocity vector $\Gamma (1, \dot{X}, \dot{Y}, \dot{Z})$ with $\dot{X} = dX/dT$, etc., in the Fermi frame can gain enough tidal energy such that its Lorentz $\Gamma$ factor could approximately approach infinity. The purpose of this paper is to extend the analysis of Ref.~\cite{Bini:2017uax} beyond the rotation axis of Kerr spacetime. 

Imagine the motion of free test particles in the exterior Kerr spacetime. The background is stationary, axisymmetric and asymptotically flat. The projections of the 4-velocity vector $u^\mu$ of a free test particle on the background Killing vector fields $\partial_{\hat{t}}$ and $\partial_{\phi}$ result in constants of the motion that are proportional to the total energy and orbital angular momentum of the free test particle as viewed by the static inertial observers at spatial infinity. Closer to the collapsed configuration, other observer families can determine the characteristics of the motion of free test particles. From a Newtonian perspective, the deeper gravitational potential near the collapsed system leads to particles with higher kinetic energies. The standard description of geodesic motion is thus anchored on the asymptotic inertial observers. On the other hand, to interpret astrophysical observation of phenomena associated with the collapsed configuration, one can instead establish a Fermi normal coordinate system around the world line of an observer that is spatially at rest near the collapsed source. Such observers in effect define the rest frame of the collapsed system. One can then study directly the gravitational influence of the collapsed system on the motion of free test particles within such a Fermi system. 

The free test particle has 4-velocity vector $u^\mu = dx^\mu/d\tau$, where $\tau$ is the proper time along its world line. Its 4-acceleration vector vanishes by definition; that is,  $Du^\mu/d\tau = 0$. To determine how the test particle's speed varies near the collapsed source, we need to refer the particle's 4-velocity vector to the tetrad frame field of a background family of observers; hence, we must deal with observer-dependent quantities~\cite{Chicone:2004rq, Chicone:2011ie}. However, if within the approximation scheme under consideration the measured speed of the free test particle approaches the speed of light---or, equivalently, the Lorentz factor of the free test particle diverges---then, we have $d\tau \to 0$, which is a significant observer-independent feature of the motion that is mainly used in the present work. In the approximate Fermi frame,  we use the expression \emph{tidal gravitational acceleration to almost the speed of light} in order to indicate exactly this observer-independent phenomenon of interest, where the Fermi frame Lorentz factor $\Gamma$ diverges.  This \emph{evidently unphysical divergence} comes about because of the approximations involved in the construction of our Fermi system. The metric in Fermi coordinates can be written as an infinite series in powers of $X$, $Y$ and $Z$; however, we ignore terms of third and higher orders in this series. Indeed, it can be shown that the inclusion of higher-order tidal terms mitigates the divergence~\cite{Chicone:2005da, Chicone:2005vn}. Nevertheless, its occurrence is an invariant indication that the particle has gained a speed close to the speed of light. In the last part of this paper (Section V), we examine timelike geodesic motion in the \emph{exact} Fermi coordinate system. However, the rest of this paper is based on the \emph{approximate} tidal equations.   

The general problem of motion of a test particle has been formulated  with respect to the Fermi normal coordinate system established 
around \emph{any} reference observer that is at rest in the exterior Kerr spacetime~\cite{Bini:2017uax, Mashhoon:2020tha}. The investigation of the geodesic trajectories in Kerr spacetime with respect to an arbitrary observer at rest thus involves the six-dimensional state space $(X, Y, Z, \dot{X}, \dot{Y}, \dot{Z})$, which is a daunting task. To simplify matters, we can set the rotation of the source equal to zero and deal instead with the spherically symmetric exterior Schwarzschild spacetime. In this case, the $X$ axis now becomes any radial direction in the exterior Schwarzschild spacetime. The absence of rotation makes it possible to set $Z = 0$, say, and the problem reduces to the four-dimensional state space $(X, Y, \dot{X}, \dot{Y})$, which is studied in the present paper. 

Throughout this work, we employ units such that $G = c = 1$, unless specified otherwise. Moreover, Greek indices run from 0 to 3, while Latin indices run from  1 to 3, and the signature of the spacetime metric is $+2$.

\section{Tidal Equations in Exterior Schwarzschild Spacetime}

Imagine a central spherical source of mass $M$ and an external test observer that is spatially at rest in the exterior Schwarzschild spacetime. Consider the radial line that connects the reference observer to the center of the spherically symmetric  gravitational source. We take this to be a radial direction of the natural $(\hat{t}, r, \theta, \phi)$ spherical polar coordinate system of the Schwarzschild gravitational field. The observer is then fixed at $r > 2 M$ and carries a natural orthonormal tetrad frame with spatial axes that are along the spherical polar coordinate directions. Using this nonrotating local frame, we set up a Fermi normal coordinate system $(T, X, Y, Z)$ centered on the fiducial observer with its proper time as the Fermi coordinate time and with $(X, Y, Z)$ along the observer's spatial tetrad frame. The observer thus permanently occupies the spatial origin of the Fermi coordinate system (i.e., $X = Y = Z = 0$). Then, the equations of motion of a free test particle in this Fermi normal coordinate system can be expressed to linear order in $X$, $Y$ and $Z$ as
\begin{align}\label{I1}
\nonumber  \frac{d^2X}{dT^2} {}&+ (A -2E\,X)\,(1-2\dot{X}^2) +A^2\,X\,(1+2\dot{X}^2)  \\   
 &-\tfrac{2}{3}\,E\, [X(\dot{Y}^2 + \dot{Z}^2) + 2\,\dot{X}(Y\dot{Y} + Z\dot{Z})]  = 0\,,
\end{align}
\begin{align}\label{I2}
\nonumber  \frac{d^2Y}{dT^2} {}&+ E\,Y\,(1-2\dot{Y}^2) -2\,[A- (A^2 + \tfrac{7}{3}\,E)\,X]\,\dot{X}\,\dot{Y} \\
& -\tfrac{2}{3}\,E\, [Y(\dot{X}^2 - 2\,\dot{Z}^2) + 5\,Z\,\dot{Y}\,\dot{Z}]=0\,,  
\end{align}
\begin{align}\label{I3}
\nonumber  \frac{d^2Z}{dT^2} {}&+ E\,Z\,(1-2\dot{Z}^2) -2\,[A- (A^2 + \tfrac{7}{3}\,E)\,X]\,\dot{X}\,\dot{Z} \\
& -\tfrac{2}{3}\,E\, [Z(\dot{X}^2 - 2\,\dot{Y}^2) + 5\,Y\,\dot{Y}\,\dot{Z}]=0\,, 
\end{align}
where $A$ and $E$ are positive constants. Let us write the unit 4-velocity vector of the test particle in the Fermi frame as $\Gamma (1, \dot{X}, \dot{Y}, \dot{Z})$; then, the Lorentz $\Gamma$ factor is given by
\begin{equation}\label{I4}
\begin{split}
  \frac{1}{\Gamma^2} = & (1+ A\,X)^2 - (\dot{X}^2 + \dot{Y}^2+ \dot{Z}^2)\\ 
 &- E\, (2\,X^2-Y^2-Z^2) +\tfrac{1}{3}\,E\,[ 2\,(Y\,\dot{Z} - Z\,\dot{Y})^2 \\
&- (X\,\dot{Z} - Z\,\dot{X})^2 - (X\,\dot{Y} - Y\,\dot{X})^2] > 0\,.
 \end{split}
\end{equation}
Indeed, the condition $\Gamma^{-2} > 0$  is the condition that these tidal equations of motion describe a timelike path in spacetime. If in the course of forward motion in time the 4-velocity vector of the test particle becomes null (i.e., $\Gamma = \infty$), then we conclude that in our approximation scheme the free test particle has been tidally accelerated to near the speed of light.  Here, $A > 0$ is the magnitude of the radial acceleration of the test observer that keeps it in place outside the attractive gravitational source and $E> 0$ has to do with the spacetime curvature that the reference observer measures, namely, 
\begin{equation}\label{I5}
A = \frac{M}{r}\,\frac{1}{(r^2 - 2 Mr)^{1/2}}\,, \qquad E  =  \frac{M}{r^3}\,.
\end{equation}
In general, $A$ and $E$ could depend upon time $T$; however, in the case under consideration the fiducial observer is spatially at rest in the background static spacetime. 
 
A subtle point that we need to emphasize is that the velocity in Fermi coordinates, $\dot{\mathbf{X}} = (\dot{X}, \dot{Y}, \dot{Z})$, is a coordinate quantity and is not invariantly defined, except at the location of the fiducial observer (i.e., $X = Y = Z = 0$), where $|\dot{\mathbf{X}}| < 1$.  We are interested in the possibility that free test particles could be tidally accelerated to near the speed of light.  If in the course of motion the timelike condition turns null (i.e., $\Gamma = \infty$), then we have an invariant measure that the test particle has indeed  been tidally accelerated up to the speed of light. 
Such a speed of light singularity would be due to the fact that our Fermi coordinates are only approximately valid. The issue has been investigated in detail in Refs.~\cite{Chicone:2005da, Chicone:2005vn}. 
Using exact Fermi coordinates in certain special spatially homogeneous spacetimes, the mitigation of such a singularity has been explicitly demonstrated~\cite{Chicone:2005vn}. 

In the spherically symmetric Schwarzschild spacetime, we have taken $X$ to be the radial coordinate out in some direction; therefore, for a given $X$, the equations of motion exhibit complete rotational symmetry in the $(Y, Z)$ system.  It is clear that for $Y = Z = 0$, radial motion is confined to the $X$ direction. Tidal motion along the rotation axis of Kerr spacetime has been studied in detail in  Refs.~\cite{Bini:2017uax, Mashhoon:2020tha}; however, in the absence of rotation the Kerr axis becomes any radial direction in Schwarzschild spacetime. We briefly describe below previous results for the sake of completeness and include certain new aspects of the tidal acceleration mechanism along the $X$ direction. Furthermore, the spherical symmetry of background spacetime makes it possible to set $Z = 0$, say, in the tidal equations. That is, starting with initial data $(X, Y, 0, \dot{X}, \dot{Y}, 0)$ in  the above system of equations, the trajectory of the free test particle remains in the $(X, Y)$ plane. In this way, we plan to extend the discussion of tidal motion to the $(X, Y)$ plane.

Complete background material regarding the issues under consideration in this paper is contained in Refs.~\cite{Bini:2017uax, Mashhoon:2020tha}.  High energy astrophysical jets are believed to be associated with compact gravitationally collapsed systems~\cite{BPu}. Astronomical observations record the motion of energetic particles relative to the central collapsed source. We therefore imagine the class of fundamental observers that are all spatially at rest in the stationary exterior of Schwarzschild spacetime  and thus represent the rest frame of the central source, which could be a black hole. We establish a Fermi normal coordinate system along the world line of such an observer. The Fermi normal coordinate system constitutes an invariantly defined geodesic reference system that is a natural extension of the inertial Cartesian frame of reference to accelerated systems and gravitational fields. It is not possible to give exact analytic expressions for the metric in such Fermi coordinates in the background Schwarzschild spacetime; indeed, only series expansions in powers of the distance away from the origin of Fermi coordinates (i.e., the position of the fiducial observer) is possible~\cite{Chicone:2005vn, BGJ}. This is in fact the case for most gravitational fields. For the sake of simplicity, we work only to linear distance away from the reference observer. 

It is clear from physical considerations in connection with the definition of Fermi coordinates that they should be admissible in a finite cylindrical spacetime domain about the path of the reference observer. The radius of this domain, $\mathcal{L}(T)$, is roughly determined by a certain infimum of the acceleration length of the fiducial observer and the measured radius of curvature of spacetime.  Reliable results are therefore expected for 
\beq \label{I5a}
|\mathbf{X}| := (X^2+Y^2+Z^2)^{1/2}  \ll \mathcal{L}\,.
\eeq
A more precise approach regarding the domain of applicability of Fermi coordinates involves the conditions for the admissibility of the spacetime coordinates. 

The standard admissibility conditions for the spacetime coordinates in a metric with $(-, +, +, +)$ signature is that the metric tensor in matrix form $(g_{\hat \mu \hat \nu})$ be negative definite,  which means that all of its principal minors must be negative~\cite{Bini:2012ht, LL}.  In the general case of a spacetime metric in Fermi coordinates, it has not been possible to express the condition of the admissibility of the Fermi coordinates in a compact form; that is, the issue must be investigated in each case separately. We are therefore interested in the boundary of the cylindrical region in spacetime along the timelike world line of the fiducial observer in  which the Fermi coordinate system is admissible. For the background exterior Schwarzschild spacetime, the components of the Fermi metric are given by~\cite{Bini:2017uax, Mashhoon:2020tha}
\beq \label{I6}
\begin{split}
&g_{\hat 0 \hat 0} = - (1 + AX)^2 + E (2 X^2 - Y^2 - Z^2) + \cdots\,, \quad \\&g_{\hat 0 \hat i} = 0 + \cdots\,,
\end{split}
\eeq
\beq \label{I7}
\begin{split}
&g_{\hat 1 \hat 1} = 1 + \tfrac{1}{3} E (Y^2+Z^2) + \cdots\,,\quad \\&g_{\hat 1 \hat 2} = -\tfrac{1}{3} E XY + \cdots\,,\quad \\&g_{\hat 1 \hat 3} = -\tfrac{1}{3} E XZ + \cdots\,
\end{split}
\eeq
and
\beq \label{I8}
\begin{split}
&g_{\hat 2 \hat 2} = 1 + \tfrac{1}{3} E (X^2 - 2\,Z^2) + \cdots\,,\quad \\&g_{\hat 2 \hat 3} = \tfrac{2}{3} E YZ + \cdots\,, \quad \\&g_{\hat 3 \hat 3} = 1 + \tfrac{1}{3} E (X^2 - 2\,Y^2) + \cdots\,.
\end{split}
\eeq

Let us note that in $g_{\hat 0 \hat 0}$,
\beq \label{I9}
2E-A^2 = E \, \frac{2r- 5M}{r- 2M}\,,
\eeq
so that there are two cases to consider: either $2M < r < (5/2) M$ or $r > (5/2)M$. We are particularly interested in the tidal acceleration mechanism (``jet effect") explored in Refs.~\cite{Bini:2017uax, Mashhoon:2020tha}. Therefore, we  will  henceforth assume that
\beq \label{I10}
2E > A^2\,, \qquad  r >  \tfrac{5}{2}\,M\,.
\eeq

We have neglected third-order terms, etc.,  in the Fermi metric; therefore, the admissibility conditions derived on the basis of this metric could only be approximately valid. If the physical region of interest is around the origin (i.e., the position of the reference observer) and  sufficiently far from the boundary of the domain of admissibility of Fermi coordinates, then the influence of the higher-order terms that have been neglected would be small and the physical results would be approximately valid.

\subsection{Admissible Region Along the $X$ Direction} 
 
If tidal motion is restricted to the X axis (i.e., the radial direction), then the standard admissibility conditions, which require that the principal minors of the matrix $(g_{\hat \mu \hat \nu})$ be all negative, reduce in this case to 
\beq \label{I11}
g_{\hat 0 \hat 0}(X, 0, 0) = - (1 + AX)^2 + 2E X^2 < 0\,,
\eeq
where we have neglected higher-order terms. This condition is simple to implement; that is,  from $(2E)^{1/2} > A$, we find that condition~\eqref{I11} is satisfied for 
\beq \label{I12}
 X_{-} < X < X_{+}\,, \qquad X_{\pm} = \frac{1}{\pm(2E)^{1/2}  - A}\,.
\eeq
Let us note that for $r \gg M$, $X_{\pm} \approx \pm (2E)^{-1/2}$.

\subsection{Admissible Region in the $(X, Y)$ Plane}

When motion is restricted to the $(X, Y)$ plane, the standard admissibility conditions reduce to 
\beq \label{I13}
g_{\hat 0 \hat 0}(X, Y, 0) = - (1 + AX)^2 + E (2X^2-Y^2) < 0\,
\eeq
and
\beq \label{I14}
 g_{\hat 3 \hat 3}(X, Y, 0) = 1+ \tfrac{1}{3} E (X^2 - 2Y^2) > 0\,.
\eeq
The first condition may be expressed as
\beq \label{I15}
\frac{(X - p)^2}{\mathbb{A}^2} - \frac{Y^2}{\mathbb{B}^2} < 1\,,
\eeq
where
\beq \label{I16}
p = \frac{A}{2E-A^2}\,, \,\,\,\,  \mathbb{A} = \frac{p}{A}\,(2E)^{1/2}\,, \,\,\,\, \mathbb{B} = \left(\frac{2p}{A}\right)^{1/2}\,.
\eeq
The second admissibility relation corresponds to the relation
\beq \label{I17}
\frac{X^2}{\mathbb{P}^2} - \frac{Y^2}{\mathbb{Q}^2} > - 1\,,
\eeq
where
\beq \label{I18}
\mathbb{P} = \sqrt{2}\,\mathbb{Q}\,, \qquad  \mathbb{Q} = \left(\frac{3}{2E}\right)^{1/2}\,.
\eeq
The allowed region is thus around the origin of Fermi coordinates in the  $(X, Y)$ plane between the two hyperbolas corresponding to the admissibility conditions~\eqref{I15} and~\eqref{I17}.  

\subsection{Dimensionless Equations}

Before we explore the consequences of the tidal equations, it proves convenient to define 
\begin{equation}\label{I19}
\begin{split}
&(t, x, y, z) := \frac{1}{r} (T, X, Y, Z)\,, \quad  a:= r\,A\,,\\&(v_x, v_y, v_z) := (\dot{X},\dot{Y}, \dot{Z})\,, \quad \quad b := r^2\,E\,.
\end{split}
\end{equation}
Moreover, $a >0$ and $b>0$ are \emph{not} independent constants; in fact
\begin{equation}\label{I19a}
\begin{split}
&a = \frac{b}{\sqrt{1-2b}}\,, \qquad\qquad\,\,\,\,  0 < a < \infty\,, \\& b = a (\sqrt{1+a^2} - a )\,, \qquad 0< b < \tfrac{1}{2}\,.
\end{split}
\end{equation}
Note that if $a = b = 0$, there is no Schwarzschild source and we are back to a free particle that moves in an inertial reference frame in flat Minkowski spacetime. 

To illustrate our dimensionless  notation, we note that the admissible region for purely radial motion~\eqref{I12} can be expressed as   
\begin{equation}\label{I20}
 x_{-} < x < x_{+}\,, \qquad x_{\pm} =\frac{1}{r}\,X_{\pm}= (\pm \sqrt{2b} -a)^{-1}\,. 
\end{equation}
It is important to note that $x_{\pm}$ are the same points in the $(x,v_x)$ plane where for $v_x=0$, we have $\Gamma = \infty$, i.e., the timelike condition is violated. Similarly, the admissible region for motion in the $(x, y)$ plane is given by
\begin{equation}\label{I21}
\begin{split}
&(1+ ax)^2-b\,(2 x^2-y^2) > 0\,,\\& 1+\tfrac{1}{3}\,b\,(x^2-2y^2) > 0\,.
\end{split}
\end{equation}

\subsection{Motion in the $x$ Direction}

It is interesting to discuss purely radial tidal motion in Schwarzschild spacetime.  With $y = z = 0$, the tidal equations reduce to  $dx/dt = v_x$ and
\begin{equation}\label{I22}
\frac{dv_x}{dt} + (a -2b\,x)\,(1-2\,v_x^2) +a^2\,x\,(1+2\,v_x^2)  = 0\,.
\end{equation}
The path is timelike if
\begin{equation}\label{I23}
\frac{1}{\Gamma^2} =  (1+ a\,x)^2 - v_x^2 -  2\,b\,x^2  > 0\,;
\end{equation}
moreover, as indicated in Eq.~\eqref{I20}, the $x$ coordinate is admissible for  $x_{-} < x < x_{+}$. The $(x, v_x)$ system has a rest point $(x_{\text{rp}}, 0)$, where
\begin{equation}\label{I24}
 x_{\text{rp}} = \frac{a}{2\,b-a^2}\,; 
\end{equation}
in this case, the timelike condition implies $2b > a^2$. This is consistent with Eq.~\eqref{I10}; that is, 
\begin{equation}\label{I24a}
 x_{\text{rp}} > 0\,,  \qquad  b = \frac{M}{r} < 0.4\,, \qquad r > 2.5\, M\,.
\end{equation}
We note that $x_{\text{rp}} = X_{\text{rp}}/r = p/r$, where $p$ is defined in Eq.~\eqref{I16}. For $r: \infty \to 2.5\,M$, $b: 0 \to 0.4$ and $x_{\text{rp}}: 0.5 \to \infty$. The saddle-type rest point  represents an unstable equilibrium~\cite{Bini:2017uax}. 

\begin{figure}
\includegraphics[width = 7cm]{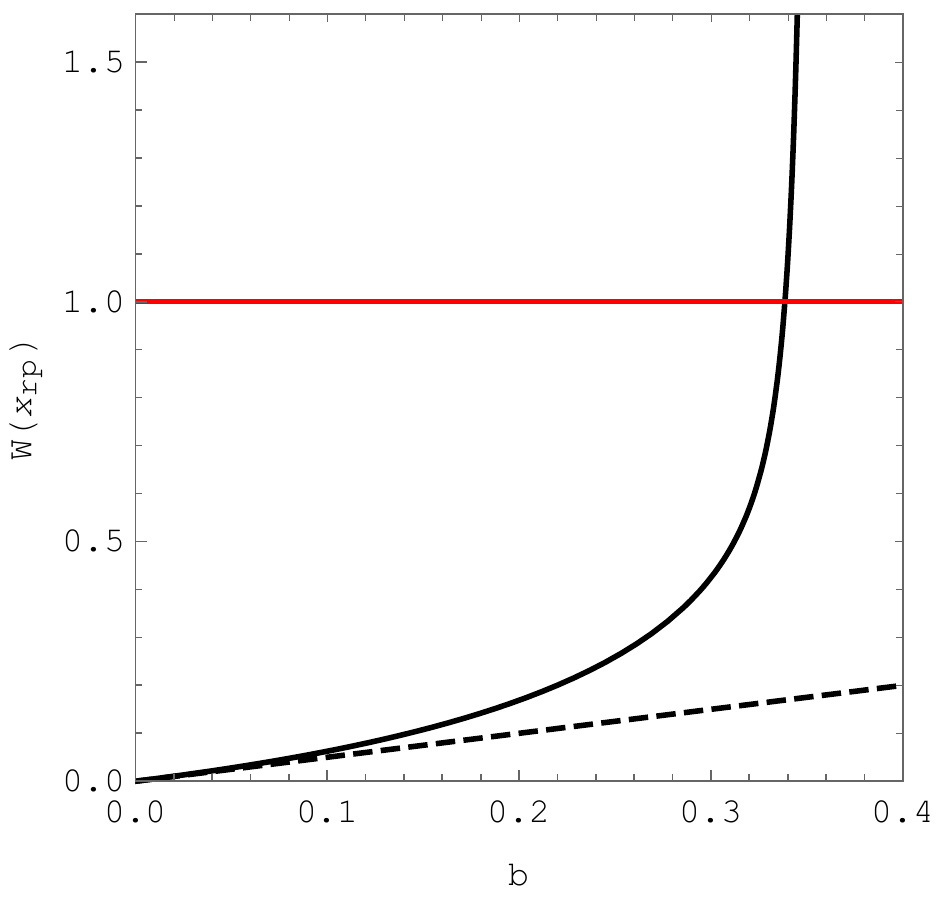}
\caption{Plot of $W(x_{\text{rp}})$ versus $b = M/r$. Note that for $b = 0.314$, $W \approx 0.54$, while for $b = 0.344$, $W \approx 1.55$. Subsequently, $W$ diverges very rapidly as $b \to 0.4$. The dashed line is $W=b/2$, which matches the solid curve for small $b$. The physical regime is limited to $W(x_{\text{rp}}) < 1$; hence, $b: 0 \to b_{\text{max}}$, $b_{\text{max}} \approx 0.34$.}
\label{wxrp}
\end{figure}

Using $dv_x /dt = v_x\,dv_x/dx$, the equation of motion~\eqref{I22} can be once integrated via an integrating factor $f(x)$ and the result is
\begin{equation}\label{I25}
f(x)\, v_x^2 + W(x) = v_x^2\mid_{x = 0} \,\, < 1\,,  
\end{equation}
which can be interpreted as an  energy equation for a classical particle in an effective potential $W(x)$. Here, 
\begin{equation}\label{I26}
f(x)= \exp{[2(2b+a^2)\,x^2-4a\,x]}>0\,
\end{equation}
and
\begin{equation}\label{I27}
\begin{split}
 &W(x) = 2\,\int_{0}^{x} [a - (2b-a^2)\xi]\,f(\xi) d\xi\,, \\&W(x) \le v_x^2\mid_{x = 0}\,.
 \end{split}
\end{equation}
The effective potential is a simple potential barrier that vanishes at the location of the fiducial observer and has a maximum at $x_{\text{rp}}$. 
Free test particles that start at $x = 0$ and move radially outward with initial $v_x^2$ less than the maximum height of the barrier, i.e., $W(x_{\text{rp}})$, have turning points and fall back toward the source, while those with initial $v_x^2 > W(x_{\text{rp}})$ clear the barrier and can eventually accelerate toward the speed of light with $\Gamma = \infty$. That is, $\Gamma^{-2}$, given by Eq.~\eqref{I23}, decreases for particles that pass over the barrier and continue along the positive $x$ axis; moreover, for some particles, $\Gamma^{-2}$ may vanish well within the boundaries of the Fermi coordinate patch~\cite{Bini:2017uax}.
The rest point $(x, v_x) = (x_{\text{rp}}, 0)$ is an exact solution of the equations of motion; therefore,  Eq.~\eqref{I25} implies that the height of the barrier is given by
\begin{equation}\label{I28}
W(x_{\text{rp}}) = v_{\text{crt}}^2 < 1\,. 
\end{equation} 
Here, $v_{\text{crt}}$ is the critical speed above which tidal acceleration toward the speed of light is possible within our approximation scheme. The barrier height as a function of $b$ is shown in Fig.~\ref{wxrp}. We find that $b: 0 \to b_{\text{max}}$, $b_{\text{max}} \approx 0.34$, corresponding to $r: r_{\text{min}} \to \,\infty$, $r_{\text{min}} \approx 2.94\, M$, since $W(x_{\text{rp}})$ must be less than unity. 

For $r \gg M$, $0 < b  \ll1$, and we find $v_{\text{crt}} \approx \sqrt{b/2}$, see the dashed line in Fig.~\ref{wxrp}. The tidal acceleration mechanism briefly described here has been investigated in Refs.~\cite{Bini:2017uax, Mashhoon:2020tha} for motion along the rotation axis of Kerr spacetime in connection with astrophysical jets; see, in particular, the phase portrait in Figure 1 of Ref.~\cite{Bini:2017uax}.

In connection with the jet effect, another phenomenon deserves attention as well. This has to do with  the trajectories of certain free test particles that could fall radially toward the gravitational source from the other side of the barrier, reverse direction after experiencing a turning point and in due course reach $\Gamma = \infty$. Indeed, as is clear from some of the trajectories in the lower right corner of Figure 1 of Ref.~\cite{Bini:2017uax}, 
free particles could fall inward and have a turning point on the right-hand side of the barrier and then move radially out and eventually accelerate to the speed of light.   
As a result of collisions in the jet plasma, there will be particles that fall down, but the turning points on the right side of the potential barrier could send them back out into the receding jet.

The tidal acceleration to the speed of light occurs along the radially outward direction within our approximation scheme, where higher-order tidal terms have been neglected. The inclusion of such terms would moderate the singularity; nevertheless, the tidal acceleration mechanism is significant and deserves further investigation. The rest of this paper is devoted to exploring this mechanism in the $(x, y)$ plane.

\section{Motion in the $(x, y)$ Plane}

Consider the motion of a free test particle in the $(x, y)$ plane. Employing the governing equations~\eqref{I1} and~\eqref{I2}, we have 
\begin{align}\label{S1}
\frac{dx}{dt} = v_x\,, \qquad \frac{dy}{dt} = v_y\,,
\end{align}
\begin{align}\label{S2}
\begin{split}
\frac{dv_x}{dt}\, +\,& (a -2b\,x)\,(1-2\,v_x^2) +a^2\,x\,(1+2\,v_x^2) 
\\& -\tfrac{2}{3}\,b\,v_y\,(x\,v_y + 2\, y\,v_x) = 0\,
 \end{split}
\end{align}
and
\begin{align}\label{S3}
\begin{split}
 \frac{dv_y}{dt} {}\,+\, &b\,y\,(1-\tfrac{2}{3}\,v_x^2-2\,v_y^2) \\&-2\,[a- (a^2 + \tfrac{7}{3}\,b)\,x]\,v_x\,v_y= 0\,,
 \end{split}
\end{align}
while the condition that the test particle follows a  timelike world line reduces to
\begin{align}\label{S4}
\begin{split}
\frac{1}{\Gamma^2} = &(1+ a\,x)^2 - (v_x^2 + v_y^2) \\&- b\, (2\,x^2-y^2) -\tfrac{1}{3}\,b\,(x\,v_y - y\,v_x)^2 > 0\,.
 \end{split}
\end{align}
Here, the physical region in the $(x, y)$ plane is the admissible region defined by Eq.~\eqref{I21}.  We wish to study the solutions of this time-invariant (autonomous) nonlinear dynamical system.

\subsection{Exact Solutions of Tidal Equations}

Given initial conditions for $x$, $y$, $v_x$ and $v_y$ at $t = 0$, say, the temporal evolution determined by the autonomous tidal equations results in a directed trajectory in the four-dimensional state space $(x, y, v_x, v_y)$. Regarding the complex nonlinear nature of the equations of motion, it is not possible to provide a complete analytical view of the solutions.  However, this system has two simple exact solutions. 

The first simple solution of the tidal equations is an isolated singularity. Indeed, our dynamical system has a rest point. To find the critical points of our system, we assume
\begin{equation}\label{S5}
\frac{dx}{dt} = \frac{dy}{dt} = 0\,,  \qquad  \frac{dv_x}{dt} = \frac{dv_y}{dt} = 0\,,
\end{equation}
Inspection of Eqs.~\eqref{S1}--\eqref{S3} reveals that we have a single rest point given by
\begin{equation}\label{S6}
 x_{\text{rp}} = \frac{a}{2\,b-a^2}\,,  \qquad  y_{\text{rp}} = 0\,,
\end{equation}
while the timelike condition is satisfied  in this case since $2\,b-a^2 > 0$. The situation here is essentially the same as when we confine our attention to the radial direction alone.   
It is remarkable that within our tidal framework a free test particle can have a \emph{rest point} in the exterior Schwarzschild spacetime; however, the rest point represents an \emph{unstable} equilibrium point as we demonstrate below. 

Let us look for the solution of tidal Eqs.~\eqref{S1}--\eqref{S3} that is infinitesimally close to the solution given by the rest point of the system~\eqref{S6}; that is, 
\begin{equation}\label{S7}
 x = \frac{a}{2\,b-a^2} +\epsilon\, \xi(t)\,,  \qquad  y = \epsilon\,\eta(t)\,,
\end{equation}
where $0<\epsilon\ll1$ is a perturbation parameter and will be treated to linear order throughout. Substitution of this solution in Eqs.~\eqref{S1} and \eqref{S2} results in 
\begin{equation}\label{S8}
 \frac{d^2\xi}{dt^2} - \lambda_1^2 \,\xi = 0\,,  \qquad  \frac{d^2\eta}{dt^2} + \lambda_2^2 \,\eta = 0\,,
\end{equation}
where
\begin{equation}\label{S9}
 \lambda_1 = (2\,b-a^2)^{1/2}\,,  \qquad  \lambda_2 = b^{1/2}\,.
\end{equation}
Moreover, the timelike condition is satisfied, as it does not change to linear order in $\epsilon$. From
\begin{equation}\label{S10}
\begin{split}
&\xi (t) = C_1 \cosh (\lambda_1\,t) + C_2 \sinh (\lambda_1\,t)\,,   \\& \eta(t) = \mathbb{C}_1 \cos (\lambda_2\,t) + \mathbb{C}_2 \sin (\lambda_2\,t)\,,
\end{split}
\end{equation}
where $(C_1, C_2)$ and $(\mathbb{C}_1, \mathbb{C}_2)$ are integration constants, we see that the equilibrium at the rest point is indeed linearly unstable due to the presence of the hyperbolic functions.  The saddle point in the case of purely radial motion similarly represents an unstable equilibrium. 

As indicated in Ref.~\cite{Mashhoon:2020tha}, 
inspection of Eqs.~\eqref{S1}--\eqref{S3} reveals the existence of a second exact solution of this system given by a \emph{rest point} $(x, v_x) = (x_0, 0)$ along the $x$ direction and a \emph{critical current} along the $y$ direction with speed $1/\sqrt{2}$. That is,  
\begin{equation}\label{S11}
 x_0 = \frac{3\,a}{7\,b-3a^2}\,, \qquad 0 < x_0 < x_{\text{rp}}\,, \qquad a \ne \frac{7}{\sqrt{51}}\,
\end{equation}
 and 
\begin{equation}\label{S12}
y = y_0 \pm \frac{\sqrt{2}}{2}\,(t-t_0)\,,
\end{equation}
where $y_0$ and $t_0$ are constants.  A linear perturbation analysis about this exact solution has been presented in Ref.~\cite{Mashhoon:2020tha}. The linearized equations were solved by the method of Frobenius, which resulted in three independent power series solutions. These indicate that the second exact solution is linearly unstable as well. No clear physical interpretation of the second exact solution has been possible thus far.

\subsection{Approximate Solutions}

It is clear from Eqs.~\eqref{S1}--\eqref{S3} that it is possible to set $y = 0$ and we then get, along the radial direction, the relatively simple ordinary differential equation~\eqref{I22}. To go further, let us assume that 
\begin{equation}\label{S13}
 y  = \theta_0 \,\zeta (t)\,, \qquad 0< \theta_0 \ll 1\,.
\end{equation}
Treating $\theta_0$ as a perturbation parameter and ignoring $\theta_0^2$ and higher-order terms, we find that our autonomous system reduces to Eq.~\eqref{I22} for $x(t)$ and a second order equation for $\zeta(t)$. Regarding the former, the last part of the previous section contains a treatment of Eq.~\eqref{I22}. The latter equation is
\begin{align}\label{S14}
 \frac{d^2 \zeta}{dt^2} + \mathcal{W}_1(t)\,\frac{d \zeta}{dt} +\mathcal{W}_2(t)\, \zeta = 0\,,  
\end{align}
where
\begin{align}\label{S15}
\mathcal{W}_1 = 2[(a^2 + \tfrac{7}{3}\,b)x(t) - a]v_x(t)\,, \quad \mathcal{W}_2 = b[1-\tfrac{2}{3}\,v_x^2(t)]\,.  
\end{align}

Using an integrating factor, we can write Eq.~\eqref{S14} in the Sturm-Liouville form
\begin{align}\label{S16}
\frac{d}{dt}\left(\mathbb{W}\, \frac{d \zeta}{dt}\right) + \mathbb{W}\,\mathcal{W}_2\, \zeta = 0\,,
\end{align}
where
\begin{align}\label{S16a}
\mathbb{W} = \exp[(a^2 + \tfrac{7}{3}\,b)\,x^2(t) - 2a\,x(t)]\,.  
\end{align}

A trivial solution of Eq.~\eqref{S16} can be obtained at the rest point when $x(t) =  x_{\text{rp}} = a/(2\,b-a^2)$. Then, $ \mathbb{W}$ is a constant that drops out and we get $\zeta(t) = \mathbb{C}_1 \cos (\lambda_2\,t) + \mathbb{C}_2 \sin (\lambda_2\,t)$ with $ \lambda_2 = \sqrt{b}$, as before; see Eqs.~\eqref{S9} and~\eqref{S10}.

To find another solution that starts from $(x, y) = (0, 0)$ with initial speed $v_0$, one can use the method of Frobenius. We find
\begin{align}\label{S17}
\nonumber x = {}&v_0 \,t - \tfrac{1}{2}\,a \,(1-2v_0^2) t^2 + \tfrac{2v_0}{3}\, \\  
&\times\,[ b\,(1-2v_0^2) - \tfrac{1}{2}\,a^2\, (5-6v_0^2)]\,t^3 + \cdots\, 
\end{align}
and
\begin{align}\label{S18}
\nonumber \zeta = {}&v_0\,t + a\,v_0^2\, t^2 - \tfrac{v_0}{3}\, \\
& \times\,[ a^2\,(1-3v_0^2) + \tfrac{1}{2}b \,(1 + 4v_0^2)]\,t^3 + \cdots\,.
\end{align}
It is simple to verify numerically the validity of these expansions with $y = \theta_0\,\zeta(t)$, where $ 0 < \theta_0 \ll 1$; that is, we have a power series solution that slightly deviates away from the radial direction. 

We mention in passing that Eq.~\eqref{S16} has a second solution $\zeta'(t)$ given by
\begin{align}\label{S19}
\zeta'(t) = \zeta(t)\,\int_{C_0}^t \frac{dt'}{\mathbb{W}(t') \,\zeta^2(t')}\,,  
\end{align}
provided $\zeta$ does not vanish in the interval $(C_0, t)$. Here, $C_0$ is an integration constant. This second solution is nonanalytic and is not of physical interest here.

It is possible to find power series solutions $x(t)$ and $y(t)$ similar to Eqs.~\eqref{S17}--\eqref{S18} for any angle $\theta_0$, $\theta_0: 0 \to \pi$. It is more interesting, however, to turn to numerical solutions. 

\section{Numerical Solutions}

\begin{figure*}
\includegraphics[width = 7.3cm]{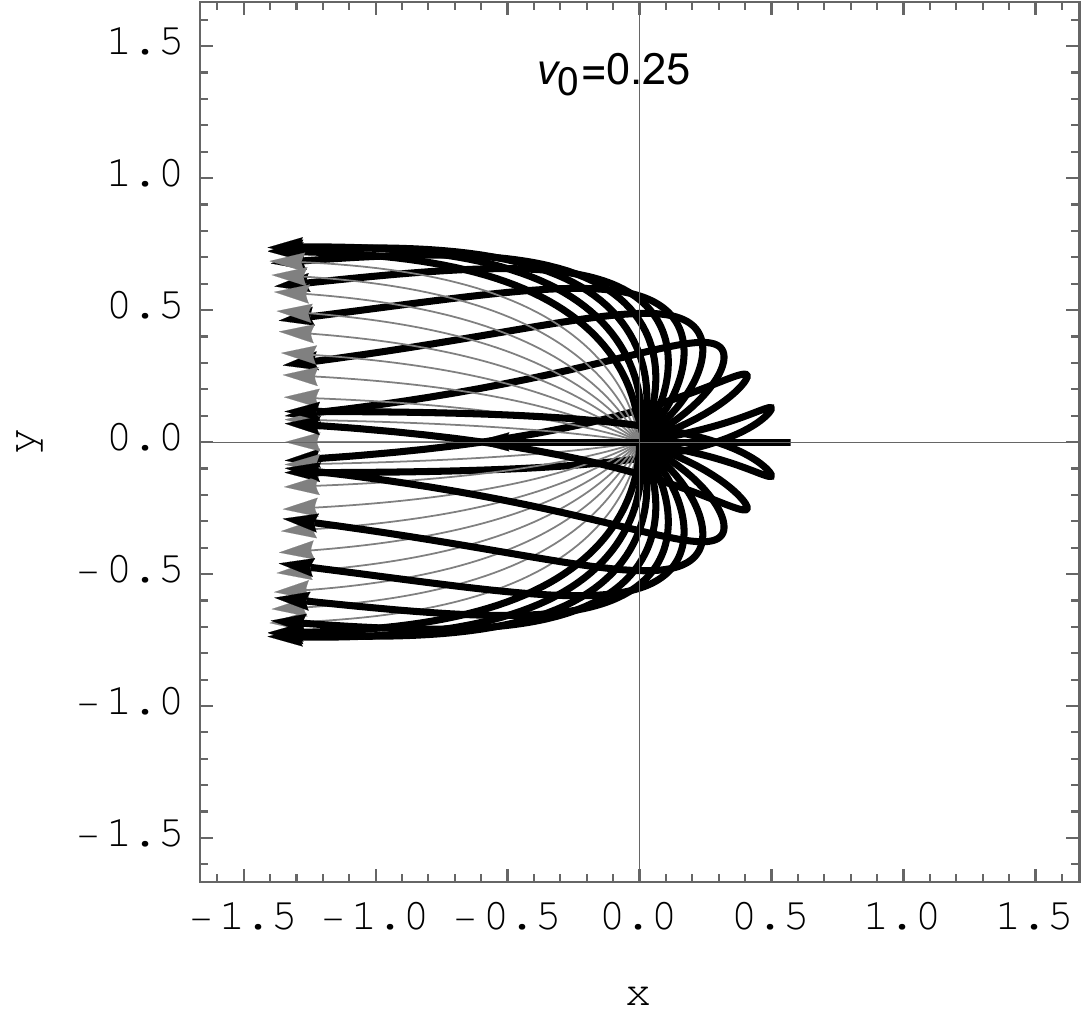}\hspace{0.2cm} \includegraphics[width = 7cm]{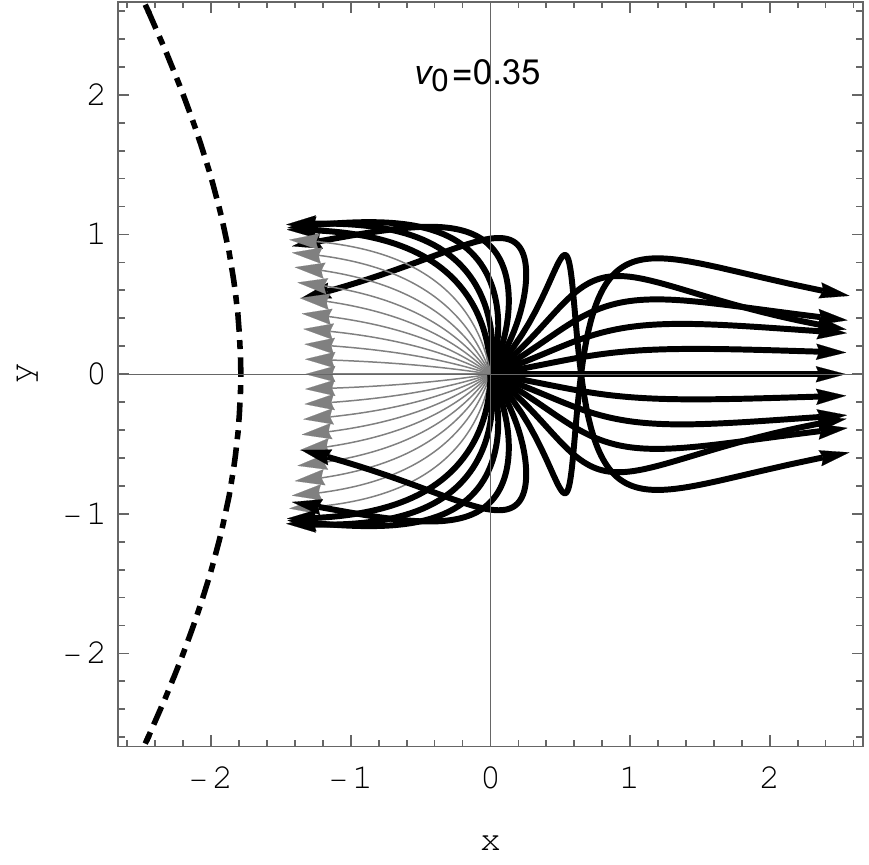}
\includegraphics[width = 7cm]{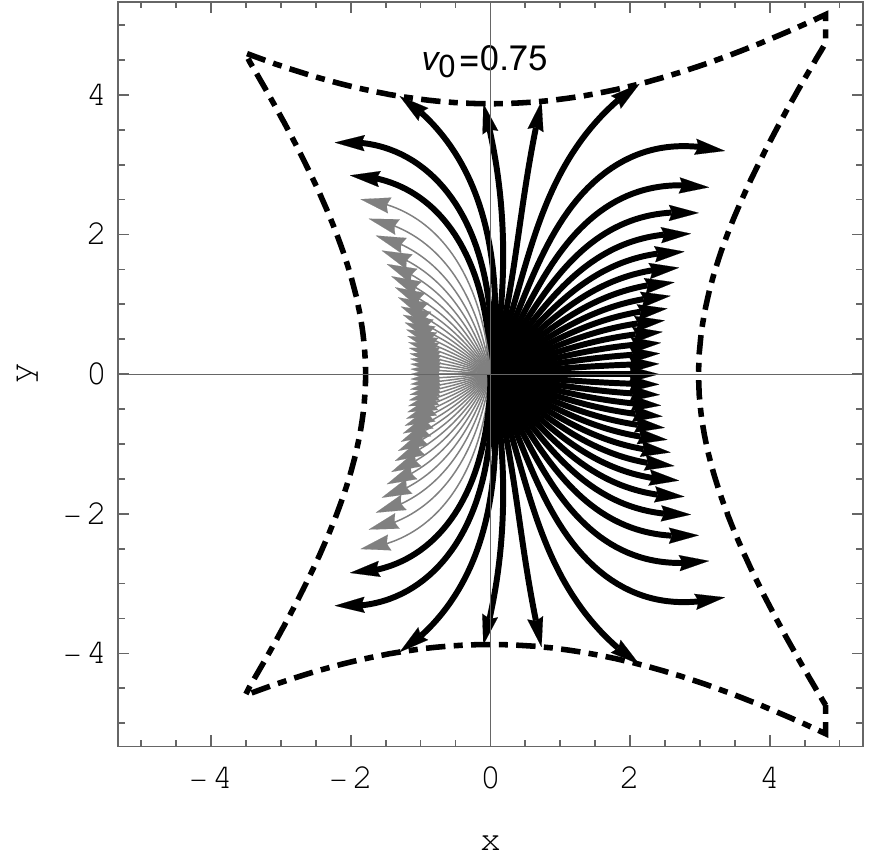}\hspace{0.2cm} \includegraphics[width = 7cm]{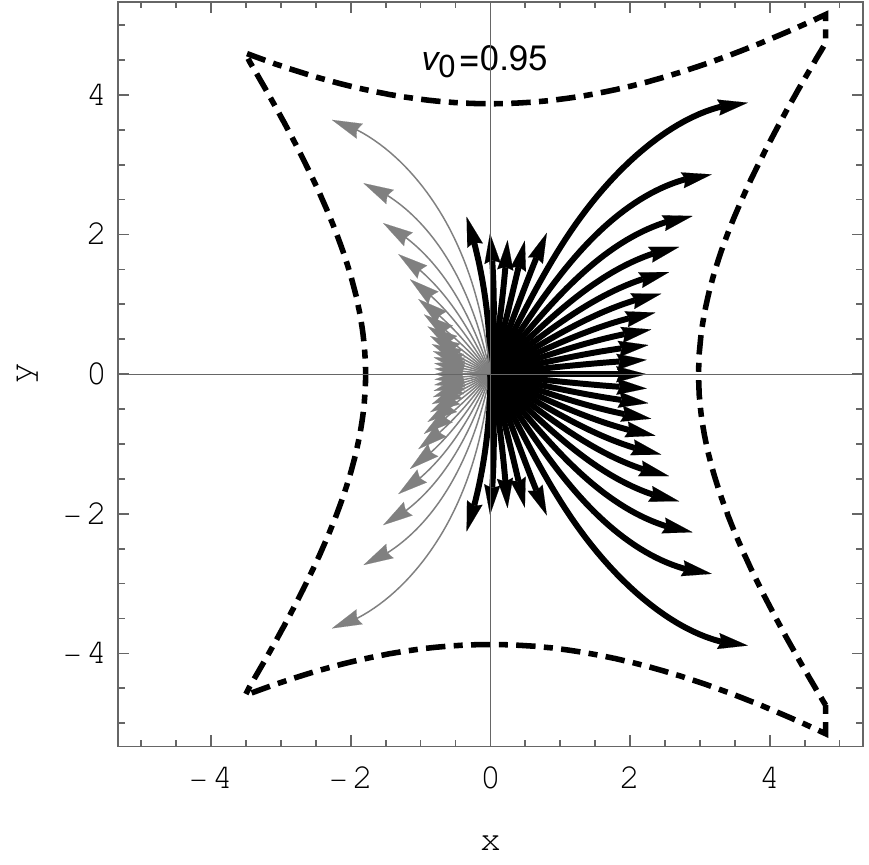}
\caption{Trajectoris of the free test particles in the $(x, y)$ plane starting from $(x,y)=(0,0)$ with a fixed velocity $v_0$ but different initial directions. The quantity $b = M/r = 0.1$ is the same in all of the panels. Dash-dotted curves indicate the boundary of the domain of admissibility of the Fermi coordinates. Integration stops when the timelike condition turns null or the test particle reaches the boundary of the Fermi coordinate patch; that is, the tip of the arrow either indicates that the test particle has been tidally accelerated to almost the speed of light or it has arrived at the boundary of the admissibility of Fermi coordinates.}
\label{nit}
\end{figure*}

The tidal equations in the $(x, y)$ plane make it possible to set $y = 0$ and restrict motion to the radial direction,  where tidal acceleration to the speed of light has already been demonstrated.  By continuity, one would also expect such tidal accelerations away from the $x$ direction as well.

To explore the behavior of the solutions of Eqs.~\eqref{S1}--\eqref{S3}, we need to choose appropriate initial conditions.
We therefore imagine free test particles that start at $t = 0$ from the origin of Fermi coordinates $(x, y) = (0, 0)$ with $v_x = v_0\,\cos\theta$ and $v_y = v_0\,\sin\theta$, where $v_0$,  $0 < v_0 < 1$, is the initial velocity and the direction of motion is arbitrary, i.e.,  $\theta: 0 \to \pi$ due to the symmetry of the underlying configuration. The results of the integrations of such trajectories are presented in Fig.~\ref{nit}. The integration stops when either the timelike condition~\eqref{S4} is violated or the particle touches the boundary of the Fermi coordinate system. In Fig.~\ref{nit}, there are four panels for different values of $v_0$. Moreover, for $\theta: 0 \to \pi/2$, the particles go out away from the source, while for $\theta: \pi/2 \to \pi$ the particles go in toward the source; in Fig.~\ref{nit}, outgoing and ingoing particles are shown by thick black curves and thin gray curves, respectively. We choose $b = M/r = 0.1$ in all of the panels. Furthermore, the dash-dotted curves denote the boundary of the Fermi coordinate patch.

In the upper left panel, $v_0=0.25$ and the gravitational attraction of the source forces all of the outgoing particles to fall back toward the source. In this case, the trajectories are all well within the admissible region. On the other hand, for $v_0=0.35$ shown in the upper right panel, the particles are more energetic and it seems that up to an angle of $\theta \approx \pi/4$, they move outward and get tidally accelerated to almost the speed of light well within the boundary of the Fermi coordinate system. In general, the strong gravitational attraction of the source leads most ingoing trajectories to reach the limit of the timelike condition quickly and thus get tidally accelerated to almost the speed of light.

In the lower left panel, we have $v_0=0.75$ and most forward-directed particles escape the gravitational source and get tidally accelerated to almost the speed of light. On the other hand, trajectories with $\theta> \pi/3$ reach the boundary of the Fermi coordinate system.

Finally, in the lower right panel, particles start with initial speed $v_0=0.95$. The remarkable result here is that particles that are moving almost perpendicularly to the radial direction reach $\Gamma = \infty$ very quickly. To shed light on this behavior, we numerically explored trajectories that started with $\theta=\pi/2$ and $0<v_0<1$. Naturally, all such particles experience backward motion toward the source of gravitational attraction. Indeed, these trajectories can be initially well approximated by
\begin{equation}\label{S20}
x(t) = -\tfrac{1}{2}\,a \,t^2 + \tfrac{a}{72}\,[15 a^2-2b(3+5v_0^2)]\,t^4  + \cdots\, 
\end{equation}
and
\begin{equation}\label{S21}
y(t) = v_0\,t - \tfrac{v_0}{6}[2a^2 + b(1-2v_0^2)]\,t^3 + \cdots\,.
\end{equation}
However, there is a critical velocity $v_{\text{crt}}\simeq 0.92$ beyond which the timelike motion turns null very quickly and well within the boundary of the Fermi coordinate patch. 

We can conclude from these numerical experiments that the acceleration phenomenon to almost the speed of light can occur in every outward direction ($\theta: 0 \to \pi/2$). This happens when the initial speed is beyond a certain threshold velocity $v_{\text{crt}}$, which for fixed $b$ increases with increasing $\theta$. Similarly, for a fixed $\theta$, $v_{\text{crt}}$ increases with increasing $b$. Let us recall that for $\theta = 0$, the particles move radially outward and there is a critical velocity that in the Schwarzschild case can be calculated using Eq.~\eqref{I28}, namely,   
\begin{equation}\label{S22}
v_{\text{crt}}=\left(\frac{b}{2}\right)^{1/2}\left(1+b+\tfrac{37}{24}\,b^2+\tfrac{659}{240}\,b^3+ \cdots\right)\,,
\end{equation}
below which the particles fall back radially into the source. For $b := GM/(c^2 r) = 0.1$, we have that $\sqrt{b/2} \approx 0.22$, so that  $v_{\text{crt}} \approx 0.2502$, which is a bit more than $v_0 = 0.25$, in agreement with the numerical results presented in the upper left panel of Fig. \ref{nit}. 

It is important to know how much of our approximate treatment survives when we employ exact Fermi coordinates in our analysis. This is the subject of the next section.

\section{Geodesic Equation in Exact Fermi Coordinate System}

The approximate Fermi coordinate system we have considered thus far is a truncated form of the exact system. Imagine the motion of a free test particle with proper time $\tau$ in the exact version of the Fermi coordinate system under consideration in this paper. The exact Fermi coordinates are $X^{\hat \mu} = (T, \mathbf{X})$, where $\mathbf{X} = (X, Y, Z)$. The reference observer is spatially at rest in the exterior (Schwarzschild or Kerr) background spacetime, which is invariant under time translation. It follows from the construction of the exact Fermi coordinate system that there is a timelike Killing vector field $\partial_T$. Indeed, the exact Fermi metric components $g_{\hat \mu \hat \nu}$, expressed as infinite series~\cite{Chicone:2005vn}, would involve constant time-independent coefficients that consist of the components of the acceleration as well as the Riemann tensor and its covariant derivatives as measured by the reference observer. Therefore, we have 
\begin{equation}\label{E1}
- d\tau^2 = g_{\hat \mu \hat \nu}(\mathbf{X})\,dX^{\hat \mu}\,dX^{\hat \nu}\,.
\end{equation}
The 4-velocity vector of a free test particle is given by  
\begin{equation}\label{E2}
U^{\hat \mu} = \frac{dX^{\hat \mu}}{d\tau} = \Gamma \,(1, \mathbf{\dot{X}}) \,, \qquad \Gamma = \frac{dT}{d\tau}\,,
\end{equation}
where an explicit expression for $\Gamma$ can be obtained from Eq.~\eqref{E1} by dividing both sides of this equation by $dT^2$. The result is
\begin{equation}\label{E3}
\Gamma = (-g_{\hat 0 \hat 0} - 2\,g_{\hat 0 \hat i}\,\dot{X}^{\hat i} - g_{\hat i \hat j}\,\dot{X}^{\hat i}\,\dot{X}^{\hat j})^{-1/2}\,.
\end{equation}
The projection of $U^{\hat \mu}$ on the timelike Killing vector field is a constant of the motion; that is,  $U\cdot \partial_T = - \mathcal{E}$. This energy integral is thus given by
\begin{equation}\label{E4}
\mathcal{E} = - g_{\hat 0 \hat \mu}\,U^{\hat \mu} = - \Gamma \,(g_{\hat 0 \hat 0} + g_{\hat 0 \hat i}\,\dot{X}^{\hat i})\,.
\end{equation}
At the position of the reference observer, which is the origin of spatial Fermi coordinates, the Fermi metric reduces to the Minkowski metric tensor. For a free test particle that starts from the position of the reference observer with initial speed $V_0$, $V_0 = |\mathbf{\dot{X}}| <1$, we find 
\begin{equation}\label{E5}
\mathcal{E} = \frac{1}{(1-V_0^2)^{1/2}}\,.
\end{equation}  

Consider, next, a class of accelerated observers that remain permanently at rest in space within the Fermi coordinate system. These observers have 4-velocity vectors given by
\begin{equation}\label{E6}
\mathbb{U}^{\hat \mu} = \frac{1}{(- g_{\hat 0 \hat 0})^{1/2}}\,\delta_{\hat 0}^{\hat \mu}\,
\end{equation}
and include the reference observer at the origin of spatial Fermi coordinates. These observers monitor the motion of the free test particles within the Fermi system. As measured by these observers, the Lorentz factor $\widehat{\Gamma}$ of a free test particle is given by
\begin{equation}\label{E7}
\widehat{\Gamma} = - g_{\hat \mu \hat \nu}\,\mathbb{U}^{\hat \mu}\,U^{\hat \nu} = \frac{\mathcal{E}}{(- g_{\hat 0 \hat 0})^{1/2}}\,.
\end{equation}
Here, $- g_{\hat 0 \hat 0}(\mathbf{X})$ must be positive by the first admissibility condition; in fact, this function is unity at the position of the reference observer, while $g_{\hat 0 \hat 0}(\mathbf{X}) = 0$ is in general a two-dimensional surface that is part of the boundary of the domain of applicability of Fermi coordinates. A free test particle that starts outward from the position of the reference observer with initial speed $V_0$ away from the (Schwarzschild or Kerr) source and arrives at the boundary of the Fermi system where $g_{\hat 0 \hat 0}(\mathbf{X}) = 0$ has a measured Lorentz factor $\widehat{\Gamma}$ that starts from $\mathcal{E}$ and eventually diverges at the boundary. Experience with the approximate Fermi system indicates that this scenario is possible only when $V_0$ is above a certain threshold escape velocity.  

For instance, when the motion of a free test particle is confined to the outgoing radial direction as in jet motion in Schwarzschild spacetime, 
\begin{equation}\label{E8}
-g_{\hat 0 \hat 0}(X) = (1+AX)^2 -2EX^2 + O(X^3)\,,
\end{equation}
which vanishes at some positive value of the $X$ coordinate  that reduces to $X_{+}$ given in Eq.~\eqref{I12} when higher-order terms are neglected. The speed of the motion is measured by the class of observers at rest in the Fermi system; indeed, this class simulates the rest frame of the (Schwarzschild or Kerr) source. As the free test particle approaches the boundary point, the measured jet speed relative to the source approaches the speed of light such that $\widehat{\Gamma}$ diverges in accordance with Eq.~\eqref{E7}. 

Similarly, in the case of motion in the $(X, Y)$ plane,  $g_{\hat 0 \hat 0}(X, Y) = 0$ in  the approximate Fermi system considered in this paper corresponds to the hyperbola whose branches cross the $X$ axis in Figure 2. Outward flow of free test particles away from the source mainly approaches the $X > 0$ branch of the hyperbola. When  free test particles reach the analogous boundary of the exact Fermi system, 
$\widehat{\Gamma}$ diverges.  As measured by the exact class of static Fermi observers, the particles are thus tidally accelerated to almost the speed of light. 

These conclusions may be compared with analogous results for the exact Fermi coordinate systems in de Sitter and G\"odel spacetimes~\cite{Chicone:2005vn}.

\section{Discussion}

We have examined the motion of  free test particles in the exterior Schwarzschild spacetime with respect to a spatially static observer that is at rest along a radial direction. This radial direction is the $X$ axis of a Fermi coordinate system with coordinates $X^{\hat \mu} = (T, X, Y, Z)$ established along the world line of the reference observer.  Working in the Fermi system, we have shown that beyond a threshold speed, tidal gravitational acceleration to almost the speed of light can take place near a collapsed configuration in certain outward directions that include the radial $X$ direction.  Within the framework of general relativity, this work contains an alternative approach to the energetics of free test particles in the vicinity of a gravitationally collapsed system. Tidal gravitational forces of the collapsed source can lead to the effective acceleration of free test particles relative to the rest frame  of the collapsed configuration.  Collisions may then lead to the dispersion of high-energy particles in every direction and produce cosmic rays; however, the presence of strong magnetic fields may confine and collimate some of these particles in the form of astrophysical jets.  The rate of accretion of matter and the magnetohydrodynamic aspects of the ambient environment clearly dominate  the outcome. To explain physical phenomena near collapsed configurations, the gravitational approach developed here must be
properly combined  with the electrodynamics of the ambient medium.

\section*{Acknowledgments}

The work of M.R. is supported by the Ferdowsi University of Mashhad. B.M. is grateful to  C. Chicone  for many helpful discussions. 

\appendix


\begin{thebibliography}{00}

\bibitem{Wheeler}
J. A. Wheeler,
``Mechanisms for Jets", 
in \emph{Nuclei of Galaxies}, edited by D. J. K. O'Connell (Elsevier, New York, 1971), pp. 539-567. 

%\cite{Blandford:1977ds}
\bibitem{Blandford:1977ds}
R.~D.~Blandford and R.~L.~Znajek,
``Electromagnetic extraction of energy from Kerr black holes,''
Mon. Not. Roy. Astron. Soc. \textbf{179}, 433-456 (1977).
%doi:10.1093/mnras/179.3.433
%2493 citations counted in INSPIRE as of 24 Nov 2020

%\cite{Blandford:1982di}
\bibitem{Blandford:1982di}
R.~D.~Blandford and D.~G.~Payne,
``Hydromagnetic flows from accretion discs and the production of radio jets",
Mon. Not. Roy. Astron. Soc. \textbf{199}, 883-903 (1982).
%1613 citations counted in INSPIRE as of 24 Nov 2020


  
\bibitem{Felice}
F. de Felice and O. Zanotti, 
``Jet Dynamics in Black Hole Physics: Acceleration During Subparsec Collimation",
Gen. Relativ. Gravit. {\bf 32}, 1449-1472 (2000).

\bibitem{Gariel:2007st} 
  J.~Gariel, M.~A.~H.~MacCallum, G.~Marcilhacy and N.~O.~Santos,
  ``Kerr Geodesics, the Penrose Process and Jet Collimation by a Black Hole'',
  Astron. Astrophys. {\bf 515}, A15 (2010)
  [gr-qc/0702123 [gr-qc]].

\bibitem{Gariel:2016vql} 
  J.~Gariel, N.~O.~Santos and A.~Wang,
  ``Observable acceleration of jets by a Kerr black hole'',
  Gen. Relativ. Gravit.  {\bf 49}, 43 (2017)
  %doi:10.1007/s10714-017-2208-9
  [arXiv:1610.01241 [gr-qc]].
  
%\cite{Zhang:2017kmq}
\bibitem{Zhang:2017kmq}
L.~Zhang, S.~B.~Chen and J.~Jing,
``Repulsive Effect for Unbound High Energy Particles Along Rotation Axis in Kerr-Taub-NUT Spacetime",
Commun. Theor. Phys. \textbf{69}, no.4, 399-406 (2018)
%doi:10.1088/0253-6102/69/4/399
[arXiv:1711.09187 [gr-qc]].
%0 citations counted in INSPIRE as of 24 Nov 2020


  
  
 %\cite{Poirier:2015cyu}
\bibitem{Poirier:2015cyu}
J.~Poirier and G.~J.~Mathews,
``Review of gravitomagnetic acceleration from accretion disks",
Mod. Phys. Lett. A \textbf{30}, no.38, 1530029 (2015).
%doi:10.1142/S0217732315300293
%0 citations counted in INSPIRE as of 24 Nov 2020

%\cite{Poirier:2016rul}
\bibitem{Poirier:2016rul}
J.~Poirier and G.~J.~Mathews,
``Gravitomagnetic acceleration from black hole accretion disks",
Classical Quantum Gravity \textbf{33}, no.10, 107001 (2016)
[arXiv:1504.02499 [gr-qc]].
%doi:10.1088/0264-9381/33/10/107001
%5 citations counted in INSPIRE as of 24 Nov 2020
  

  
 \bibitem{Tucker:2016wvt} 
  R.~W.~Tucker and T.~J.~Walton,
  ``On Gravitational Chirality as the Genesis of Astrophysical Jets'',
  Classical Quantum Gravity  {\bf 34},  035005 (2017)
  %doi:10.1088/1361-6382/aa5325
  [arXiv:1609.07322 [gr-qc]].
  
 %\cite{Tucker:2018xle}
\bibitem{Tucker:2018xle}
R.~W.~Tucker and T.~J.~Walton,
``Chirality in Gravitational and Electromagnetic Interactions with Matter",
Int. J. Geom. Meth. Mod. Phys. \textbf{15}, no.supp01, 1840004 (2018)
%doi:10.1142/S0219887818400042
[arXiv:1805.08825 [gr-qc]].
%1 citations counted in INSPIRE as of 24 Nov 2020


\bibitem{mas77}
B. Mashhoon,
``Tidal radiation'',
Astrophys. J.\ {\bf 216}, 591-609 (1977).

\bibitem{Marck}
J.-A. Marck, ``Solution to the equations of parallel transport in Kerr geometry; tidal tensor",
Proc. R. Soc. London A {\bf 385}, 431 (1983). 



\bibitem{MaMc}
B.~Mashhoon and J.~C.~McClune, ``Relativistic tidal impulse",
Mon. Not. Roy. Astron. Soc. {\bf 262}, 881-888 (1993).

%\cite{Chicone:2002kb}
\bibitem{Chicone:2002kb}
C.~Chicone and B.~Mashhoon,
``The generalized Jacobi equation'',
Classical Quantum Gravity \textbf{19}, 4231-4248 (2002)
%doi:10.1088/0264-9381/19/16/301
[arXiv:gr-qc/0203073 [gr-qc]].
%43 citations counted in INSPIRE as of 29 May 2020





%\cite{Kojima:2005dm}
\bibitem{Kojima:2005dm}
Y.~Kojima and K.~Takami,
``Tidal effects on magnetic gyration of a charged particle in Fermi coordinates'',
Classical Quantum Gravity \textbf{23}, 609-616 (2006)
%doi:10.1088/0264-9381/23/3/004
[arXiv:gr-qc/0509084 [gr-qc]].
%9 citations counted in INSPIRE as of 29 May 2020

\bibitem{Bini:2007zzb} 
  D.~Bini, F.~de Felice and A.~Geralico,
 ``Strains and axial outflows in the field of a rotating black hole'',
  Phys.\ Rev.\ D {\bf 76}, 047502 (2007)
  %doi:10.1103/PhysRevD.76.047502
  [arXiv:1408.4592 [gr-qc]].

%\cite{Bini:2012zze}
\bibitem{Bini:2012zze}
D.~Bini, K.~Boshkayev and A.~Geralico,
``Tidal indicators in the spacetime of a rotating deformed mass",
Classical Quantum Gravity \textbf{29}, 145003 (2012)
%doi:10.1088/0264-9381/29/14/145003
[arXiv:1306.4803 [gr-qc]].
%10 citations counted in INSPIRE as of 29 Nov 2020

%\cite{Bini:2017uax}
\bibitem{Bini:2017uax}
D.~Bini, C.~Chicone and B.~Mashhoon,
``Relativistic tidal acceleration of astrophysical jets'',
Phys. Rev. D \textbf{95}, no.10, 104029 (2017) 
%doi:10.1103/PhysRevD.95.104029
[arXiv:1703.10843 [gr-qc]].
%10 citations counted in INSPIRE as of 19 Apr 2020






%\cite{Junior:2020yxg}
\bibitem{Junior:2020yxg}
H.~C.~D.~Lima~Junior, L.~C.~B.~Crispino and A.~Higuchi,
``On-axis tidal forces in Kerr spacetime'',
Eur. Phys. J. Plus \textbf{135}, no.3, 334 (2020) 
%doi:10.1140/epjp/s13360-020-00342-7
[arXiv:2003.09506 [gr-qc]].
%0 citations counted in INSPIRE as of 19 Apr 2020

%\cite{Junior:2020par}
\bibitem{Junior:2020par}
H.~C.~D.~Lima~Junior and L.~C.~B.~Crispino,
``Tidal forces in the charged Hayward black hole spacetime'',
Int. J. Mod. Phys. D (2020)
[arXiv:2005.13029 [gr-qc]].
%0 citations counted in INSPIRE as of 29 May 2020

%\cite{Mashhoon:2020tha}
\bibitem{Mashhoon:2020tha}
B.~Mashhoon,
``Critical Tidal Currents in General Relativity",
Universe \textbf{6}, no.8, 104 (2020)
%doi:10.3390/universe6080104
[arXiv:2007.12023 [gr-qc]].
%0 citations counted in INSPIRE as of 25 Nov 2020






 


\bibitem{Chandra}
S.~Chandrasekhar, \emph{The Mathematical Theory of Black Holes} (Clarendon, Oxford, 1983).
  
\bibitem{Synge}
J.~L.~Synge, \emph{Relativity: The General Theory} (North-Holland, Amsterdam, 1971).  



%\cite{Chicone:2004rq}
\bibitem{Chicone:2004rq}
C.~Chicone and B.~Mashhoon,
``Tidal acceleration of ultrarelativistic particles'',
Astron. \&  Astrophys. \textbf{437}, L39-L42 (2005)
%doi:10.1051/0004-6361:200500137
[arXiv:astro-ph/0406005 [astro-ph]].
%16 citations counted in INSPIRE as of 29 May 2020

%\cite{Chicone:2011ie}
\bibitem{Chicone:2011ie}
C.~Chicone, B.~Mashhoon and K.~Rosquist,
``Double-Kasner Spacetime: Peculiar Velocities and Cosmic Jets",
Phys. Rev. D \textbf{83}, 124029 (2011)
%doi:10.1103/PhysRevD.83.124029
[arXiv:1104.5058 [gr-qc]].
%7 citations counted in INSPIRE as of 03 Jan 2021


%\cite{Chicone:2005da}
\bibitem{Chicone:2005da}
C.~Chicone and B.~Mashhoon,
``A gravitational mechanism for the acceleration of ultrarelativistic particles,''
Ann. Phys. (Berlin) \textbf{14}, 751-763 (2005)
%doi:10.1002/andp.200510164
[arXiv:astro-ph/0502560 [astro-ph]].
%6 citations counted in INSPIRE as of 29 May 2020

%\cite{Chicone:2005vn}
\bibitem{Chicone:2005vn}
C.~Chicone and B.~Mashhoon,
``Explicit Fermi coordinates and tidal dynamics in de Sitter and G\"odel spacetimes'',
Phys. Rev. D \textbf{74}, 064019 (2006)
%doi:10.1103/PhysRevD.74.064019
[arXiv:gr-qc/0511129 [gr-qc]].
%50 citations counted in INSPIRE as of 29 May 2020


\bibitem{BPu}
B. Punsly, \emph{Black Hole Gravitohydromagnetics}, 2nd ed. (Springer-Verlag, Berlin, 2008).

\bibitem{BGJ}  
D.~Bini, A.~Geralico and R.~T.~Jantzen,  
``Fermi coordinates in Schwarzschild spacetime: closed form expressions", 
Gen.\ Relativ.\ Gravit. {\bf 43}, 1837-1853 (2011) 
 [arXiv:1408.4947 [gr-qc]].
%https://doi.org/10.1007/s10714-011-1163-0











%\cite{Bini:2012ht}
\bibitem{Bini:2012ht}
D.~Bini, C.~Chicone and B.~Mashhoon,
``Spacetime Splitting, Admissible Coordinates and Causality'',
Phys. Rev. D \textbf{85}, 104020 (2012)
%doi:10.1103/PhysRevD.85.104020
[arXiv:1203.3454 [gr-qc]].
%10 citations counted in INSPIRE as of 28 Sep 2020

\bibitem{LL}
L. D. Landau and E. M. Lifshitz, 
{\it The Classical Theory of Fields}
(Pergamon, New York, 1975). 





\end{thebibliography}
\end{document}